\documentclass[aps,prc,twocolumn,superscriptaddress]{revtex4-1}

\usepackage{amsmath}
\usepackage{graphicx}

\begin{document}


\title{Measurement of $^{58}$Ni($p$, $p$)$^{58}$Ni elastic scattering at low momentum transfer by using the HIRFL-CSR heavy-ion storage ring}



\author{K. Yue}
\affiliation{Institute of Modern Physics, Chinese Academy of Science, Lanzhou 730000, China}
\author{J. T. Zhang}
\affiliation{Institute of Modern Physics, Chinese Academy of Science, Lanzhou 730000, China}
\affiliation{School of Nuclear Science and Technology, Lanzhou University, Lanzhou 730000, China}
\author{X. L. Tu}
\email{tuxiaolin@impcas.ac.cn}
\affiliation{Institute of Modern Physics, Chinese Academy of Science, Lanzhou 730000, China}
\affiliation{School of Nuclear Science and Technology, University of Chinese Academy of Sciences, Beijing 100049, China}
\affiliation{Max-Planck-Institut  f\"ur Kernphysik, Saupfercheckweg 1, 69117 Heidelberg, Germany}
\author{C. J. Shao}
\author{H. X. Li}
\author{P. Ma}
\affiliation{Institute of Modern Physics, Chinese Academy of Science, Lanzhou 730000, China}
\author{B. Mei}
\affiliation{Institute of Modern Physics, Chinese Academy of Science, Lanzhou 730000, China}
\affiliation{Sino-French Institute of Nuclear Engineering and Technology, Sun Yat-sen University, Zhuhai 519082, China}
\author{X. C. Chen}
\author{Y. Y. Yang}
\affiliation{Institute of Modern Physics, Chinese Academy of Science, Lanzhou 730000, China}
\author{X. Q. Liu}
\affiliation{Sichuan University, Chengdu 610065,  China}
\author{Y. M. Xing}
\affiliation{Institute of Modern Physics, Chinese Academy of Science, Lanzhou 730000, China}
\affiliation{GSI Helmholtzzentrum f\"ur Schwerionenforschung, D-64291 Darmstadt, Germany}
\author{K. H. Fang}
\affiliation{School of Nuclear Science and Technology, Lanzhou University, Lanzhou 730000, China}
\author{X. H. Li}
\affiliation{University of South China, Hengyang 421001, China}
\author{Z. Y. Sun}
\author{M. Wang}
\affiliation{Institute of Modern Physics, Chinese Academy of Science, Lanzhou 730000, China}
\author{P. Egelhof}
\author{Yu.  A. Litvinov}
\affiliation{GSI Helmholtzzentrum f\"ur Schwerionenforschung, D-64291 Darmstadt, Germany}
\author{K. Blaum}
\affiliation{Max-Planck-Institut  f\"ur Kernphysik, Saupfercheckweg 1, 69117 Heidelberg, Germany}
\author{Y. H. Zhang}
\author{X. H. Zhou}
\affiliation{Institute of Modern Physics, Chinese Academy of Science, Lanzhou 730000, China}


\date{\today}

\begin{abstract}
The very first in-ring reaction experiment at the HIRFL-CSR heavy-ion storage ring, namely proton elastic scattering on stable $^{58}$Ni nuclei, is presented.
The circulating $^{58}$Ni$^{19+}$ ions with an energy of 95 MeV/u were interacting repeatedly with an internal hydrogen gas target in the CSRe experimental ring.
Low energy proton recoils from the elastic collisions were measured with an ultra-high vacuum compatible silicon-strip detector. Deduced differential cross sections were normalized by measuring K-shell X-rays from $^{58}$Ni$^{19+}$ projectiles due to the $^{58}$Ni$^{19+}$-H$_2$ ionization collisions. Compared to the experimental cross sections, a good agreement has been achieved with the theoretical predictions in the measured region, which were obtained by using the global phenomenological optical model potentials. Our results enable new research opportunities for optical model potential studies on exotic nuclides by using the in-ring reaction setup at the HIRFL-CSR facility.

\end{abstract}


\maketitle

\section{INTRODUCTION}

The investigation of direct reactions induced by light ions, e.g. proton and alpha particles, provides important information on nuclear structure and astrophysics~\cite{Glendenning04}.
Elastic scattering of light ions, since the Geiger-Marsden experiment in 1908~\cite{Geiger09}, has been used widely, not only to study fundamental properties of nuclei, such as nuclear matter distributions~\cite{Egelhof02}, but also to extract optical model potentials (OMPs)~\cite{Becchetti69}, which are essential for the description of direct reactions on exotic nuclei with the distorted-wave Born approximation (DWBA)~\cite{Glendenning04}.

The proton elastic scattering on stable nuclides has been investigated both theoretically and experimentally~\cite{Sakaguchi17}. The phenomenological and microscopic OMPs were developed to understand and predict reaction cross sections. With scattering data from stable nuclides, the global phenomenological OMP parameters for proton elastic scattering on heavy ions in the energy region up to 200 MeV have been extracted theoretically~\cite{Koning03, Li08}. However, since it became clear that the radial shape and strength of the OMPs depend strongly on the proton-neutron asymmetry~\cite{Dickhoff19,Gutbrod06}, the OMPs extracted from stable nuclei can not directly be used to predict the scattering cross sections on unstable nuclei.

Direct reactions induced by light ions were mostly performed in direct kinematics, where the light-ion beams interact with a target made of the nuclei of interest~\cite{Li07}.
Obviously, such kind of experiments are limited to stable or very long lived nuclides. To explore direct reactions on exotic nuclei, experimental methods based on inverse kinematics, such as active gas targets~\cite{Egelhof02} and $^6$Li scattering~\cite{Chen09}, have been developed. In recent years, a new experimental method, namely, studies with stored beams in storage rings interacting with internal gas-jet targets has attracted much interest. The EXL (EXotic nuclei studied in Light-ion induced reactions at storage rings) project~\cite{Gutbrod06,Egelhof03} has been developed to study nuclear matter distributions~\cite{Zamora17,Schmid15,Egelhof15}, giant resonances~\cite{Zamora16} and astrophysical reaction rates~\cite{Mei15,Glorius19}, at the GSI and later FAIR facilities~\cite{Gutbrod06}. It has been demonstrated that direct reactions induced by light ions, especially for scattering processes at very low momentum transfer, which play an important role in studies of isoscalar giant monopole resonances and nuclear matter distributions,  can be successfully approached with the novel in-ring reaction experimental methods~\cite{Zamora17, Zamora16,Streicher11,Mutterer15}. For proton elastic scattering at low momentum transfer,  the cross sections are rather high. Most important is that the differential cross sections in the low momentum transfer region are very sensitive to deduce nuclear matter distributions, thus the size and radial shape of nuclei can be determined precisely~\cite{Alkhazov97,Alkhazov02,Dobrovolsky06}.

As one of the existing facilities, the Cooler Storage Ring at the Heavy Ion Research Facility in Lanzhou (HIRFL-CSR)~\cite{Xia02} provides an opportunity for performing in-ring reaction experiments by using internal gas-jet targets to study, e.g., proton elastic scattering on nuclei at low momentum transfer. Proton scattering on stable $^{58}$Ni nuclei has been measured widely at different energies, see Ref.~\cite{Koning03} and references cited therein, however, data on cross sections at low momentum transfer are still scarce. Furthermore, for the 100 MeV proton elastic scattering on $^{58}$Ni nuclei in Ref.~\cite{Kwiatkowski78}, the authors found that the differential cross sections would not be described by the optical model with a good $\chi$$^2/N$, and a search for physical reasons was proposed~\cite{Kwiatkowski78}. The inconsistency was also observed for the global phenomenological OMPs in Ref.~\cite{Koning03}. In this work, a first in-ring reaction experiment on proton elastic scattering of 95 MeV/u $^{58}$Ni$^{19+}$ ions at low momentum transfer was conducted at the HIRFL-CSR heavy-ion storage ring. The measured differential cross sections can be used to check the inconsistency reported in Refs.~\cite{Kwiatkowski78, Koning03}. The present successful experiment enables the capability for OMPs studies of exotic nuclei at the HIRFL-CSR heavy-ion storage ring.

\section{EXPERIMENT}

The experiment was carried out in inverse kinematics at the experimental storage ring CSRe of HIRFL-CSR~\cite{Xia02}. The $^{58}$Ni$^{19+}$ beam was accelerated up to an energy of 95 MeV/u by the heavy-ion synchrotron CSRm, then extracted and transported to the CSRe via the second Radioactive Ion Beam Line in Lanzhou (RIBLL2). In general, radioactive ion beams can be produced by using projectile fragmentation reactions at the RIBLL2, as discussed in, e.g., Refs.~\cite{Tu11, Zhang12}. The 95 MeV/u $^{58}$Ni$^{19+}$ ions were stored in the CSRe with an intensity of about 10$^7$ particles in each measurement cycle. The electron cooling has been applied to reduce the emittance and velocity spread of the beam~\cite{Mao16}. The cooled $^{58}$Ni$^{19+}$ beam interacted with an internal hydrogen gas-jet target oriented perpendicular to the beam direction. The internal gas-jet target has been previously applied for atomic physic studies at the CSRe~\cite{Shao17}.  A typical diameter of the gas-jet target is about 4 mm at the interaction point. A target density of about 10$^{12}$ atoms/cm$^2$ has been achieved~\cite{Shao13}.

\begin{figure}
\begin{center}
\includegraphics*[width=0.5\textwidth]{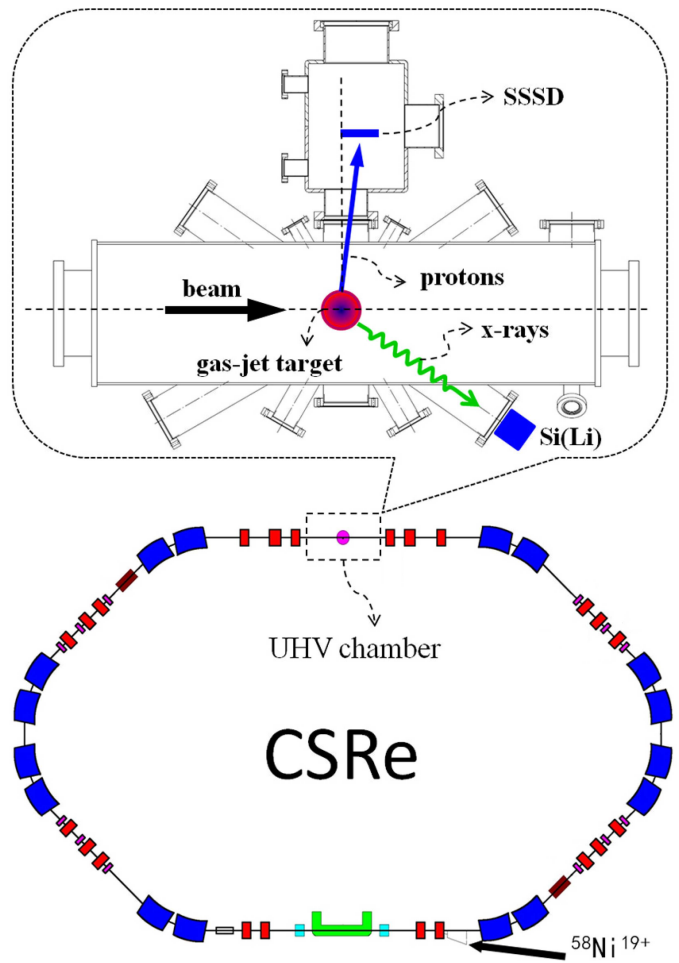}
\caption{(Color online). Schematic illustration of the detector system installed for in-ring reaction experiments. The lower part shows the CSRe experimental ring. The SSSD was installed into a pocket at an angle of  90$^\circ$ for proton elastic scattering measurements. The Si(Li) luminosity monitor was mounted in a pocket at an angle of 35$^\circ$ for measuring X-rays.}
\label{fig1}
\end{center}
\end{figure}
\begin{figure}[h!]
\begin{center}
\includegraphics*[width=0.5\textwidth]{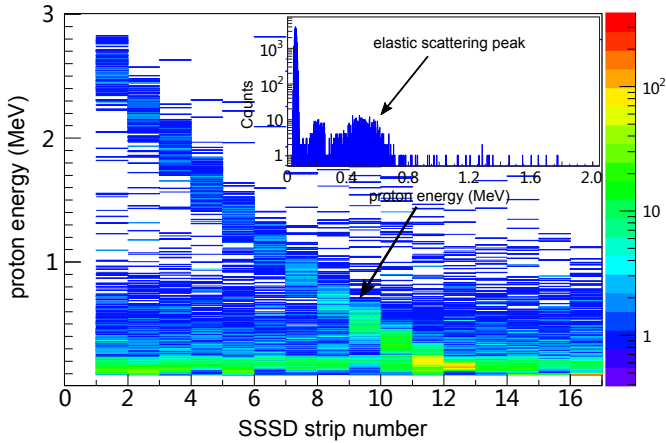}
\caption{(Color online). The proton energies of recoils measured as a function of the Si-strip number for proton elastic scattering on $^{58}$Ni nuclei at 95 MeV/u. The inset shows the energy spectrum measured by the ninth strip of the SSSD.
The SSSD with a thickness of 300 $\mu$m is thick enough to effectively stop the scattered protons in the present experiment.}
\label{fig2}
\end{center}
\end{figure}

In order to measure very low energy recoil protons, a single-sided silicon detector (SSSD) with a thickness of 300 $\mu$m has been installed in the ultra-high vacuum (UHV) chamber, which is connected to the CSRe. The SSSD is fully compatible with the UHV environment. For more details on the SSSD see Ref.~\cite{Zhang19}.  Figure~\ref{fig1} shows the schematic drawing of the experimental setup. The SSSD with an active area of 48$\times$48 mm$^2$, mounted at a distance of 503 mm from the reaction collision point, covers the laboratory angular range from 85$^\circ$ to 90$^\circ$ for proton recoils. The signals from the SSSD were fed to the Mesytec MPR-16 preamplifier. The MSCF-16 shaping amplifier was used to process the signals from the preamplifier. Afterwards, all signals were recorded by the data acquisition system (DAQ). The DAQ was triggered by a logic OR of signals from the SSSD.
A typical energy spectrum of measured proton recoils is shown in Fig.~\ref{fig2}. It was calibrated by using $^{207}$Bi, $^{239}$Pu and $^{241}$Am radioactive sources.
According to the kinematic calculation, only elastic scattering protons can be detected in the covered laboratory angular range in the absence of inelastic scattering events.

\begin{figure}[h!]
\begin{center}
\includegraphics*[width=0.5\textwidth]{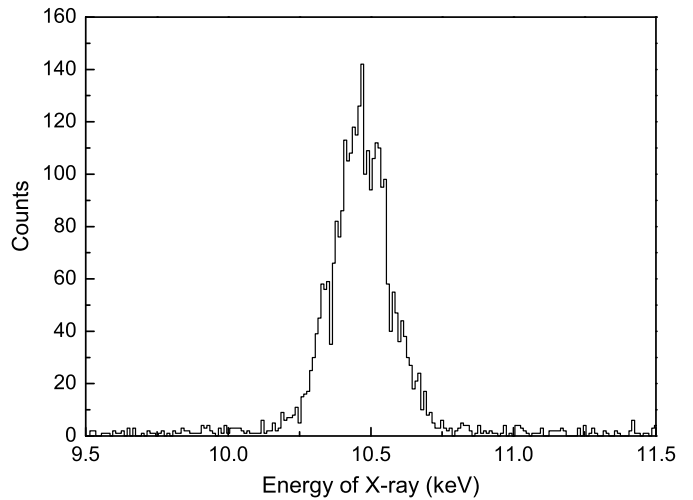}
\caption{A typical X-ray spectrum from the K-shell ionization of $^{58}$Ni$^{19+}$ ions, which was measured by the Si(Li) detector at an observation angle of 35$^\circ$. The Doppler shift is not corrected.}
\label{fig3}
\end{center}
\end{figure}

The knowledge of the reaction luminosity is essential for determining absolute cross sections. It is not easy to determine the luminosity for in-ring reaction measurements, not only due to the changes of the beam intensity and gas-target density in time, but also due to the uncertainty on the overlap between beam and gas-target. To determine accurately the reaction luminosity, the K-shell X-rays from inner-shell ionization of $^{58}$Ni$^{19+}$ ions, which were produced in the collisions with the H$_2$ target, have been measured simultaneously with a Si(Li) detector. As shown in Fig.~\ref{fig1}, the Si(Li) detector is placed at 35$^\circ$, at a distance to the collision point of 488 mm. The detector was separated from the UHV environment of the CSRe by a 100 $\mu$m beryllium window and collimated by a hole of 4 $\times$ 8 mm$^2$. The Si(Li) detector was calibrated with $^{55}$Fe, $^{109}$Cd $^{133}$Ba, and $^{241}$Am radioactive sources. A typical K-shell X-ray energy spectrum obtained in the experiment is shown in Fig~\ref{fig3}. 
The cross sections for the X-ray emissions have been extensively studied both theoretically and experimentally in atomic physics, and can be calculated with high precision~\cite{Eichler07}. Combined with the detection efficiency of the Si(Li) detector, the absolute luminosity for in-ring reaction measurements can be obtained. A Similar method has been used to determine the luminosity for in-ring reaction experiments on bare nuclei in Refs.~\cite{Mei15,Glorius19}.

\section{DATA ANALYSIS AND RESULTS}


According to the two-body kinematics in the inverse framework~\cite{Norbury08}, the SSSD with a thickness of 300 $\mu$m is thick enough to effectively stop the elastic scattered protons. The maximum measured energy of proton recoils is about 2.8 MeV, see Fig.~\ref{fig2}. The proton scattering angle in the laboratory (LAB) frame ($\theta_{p}^{LAB}$) can be determined through measuring the proton kinetic energy ($K_{p}^{LAB}$), by using the following equation~\cite{Norbury08}.
\begin{equation}\label{e1}
  2m_{p}K_{p}^{LAB}=4(p^{CM})^2(\frac{1}{1+(\gamma^{CM})^2\tan^2\theta_p^{LAB}})\quad,
\end{equation}
in which, $m_{p}$, $p^{CM}$, and $\gamma^{CM}$ are the rest mass of the proton, the momentum in the center-of-mass (CM) frame, and the Lorentz factor of the CM frame relative to the LAB frame, respectively. In this experiment, $p^{CM}$ and $\gamma^{CM}$  were 0.426 GeV, and 1.098, respectively. It is convenient to make use of the Mandelstam variable $-t$ to extract the differential cross sections for proton scattering~\cite{Norbury08}.
\begin{equation}\label{e2}
  -t=2m_{p}K_{p}^{LAB}=2(p^{CM})^2(1-\cos\theta^{CM})\quad,
\end{equation}
\begin{equation}\label{e3}
\frac{d\sigma}{dt}=\frac{1}{L}\frac{\Delta N_t}{\Delta t}\quad,
\end{equation}
where $-t$ is defined as the square of the four-momentum transfer, which can be expressed in terms of the proton kinetic energy ($K_{p}^{LAB}$) after collision with heavy ions.
$L$ is the integrated luminosity.  $\theta^{CM}$ is the scattering angle of the proton in the CM frame. $\Delta N_t$ is the number of protons in the bin size $\Delta t$. According to Eqs. (\ref{e2}) and (\ref{e3}), for elastic scattering, $\theta^{CM}$ and $-t$ can be determined by measuring the kinetic energy of the proton. Since the used SSSD has an energy resolution of better than 1$\%$~\cite{Zhang19}, this can be done with high accuracy. Thus, one can also deduce the cross sections as a function of $-t$ by using the acquired data from the SSSD. 
According to the position (angle) relations between the Si strips and the hydrogen gas target, the elastic scattering energy peak from the hydrogen gas target can be identified from the background events at each strip, especially for the measured proton peaks with energies $\textgreater$ 370 keV, see the inset in Fig.~\ref{fig2}. The backgrounds were mainly the scattering events of diffusion hydrogen gas. In order to reduce background effects, only protons in the elastic scattering peaks with $K_{p}^{LAB}$$\textgreater$ 370 keV were adopted in the present work. However, due to the probability distribution, the background events close to the tails of the elastic scattering peaks may be included, but the effects are only around 1$\%$.

The differential cross sections can also be expressed as a function of the CM angle ($\theta^{CM}$), by using~\cite{Norbury08},
\begin{equation}\label{e4}
  (\frac{d\sigma}{d\Omega})^{CM}=\frac{(p^{CM})^2}{\pi}\frac{d\sigma}{dt}\quad.
\end{equation}
As we know, the differential cross sections for proton elastic scattering on heavy ions have been investigated for over 100 years.
The OMPs have been widely used to describe the differential cross sections at low and intermediate energies. 
The OMP parameters for proton elastic scattering on stable nuclei in the energy region up to 200 MeV have been extensively studied~\cite{Koning03, Li08}.
In the following we employ the OMP parameters from  A. J. Koning and J. P. Delaroche (KD03)~\cite{Koning03} and X. Li and C. Cai (LC08)~\cite{Li08}. The differential cross sections were calculated with the coupled-reaction channels program FRESCO \cite{Thompson}. On average, an agreement between the global OMP predictions and the previous experimental results can be achieved within 10$\%$~\cite{Koning03}.

To obtain absolute differential cross sections of elastic scattering in this analysis, the luminosity was deduced by using the measured K-shell X-rays.
\begin{equation}
L=\frac{N_K}{\sigma_{K}\omega_K\varepsilon\frac{\Omega}{4\pi}}\gamma^2(1-\beta \cos\theta_{lab})^2\quad,
\label{e5}
\end{equation}
where $\Omega$ and $\theta_{lab}$ are the solid angle, and measuring angle of Si(Li) detector, respectively. A detection efficiency ($\varepsilon$) of 100$\%$ can be achieved for X-rays with an energy of 10 keV by using the Si(Li) detector~\cite{ORTEC}. The solid angle of the effective area for the Si(Li) detector is 0.134(2) msr, which was obtained by a Geant4 simulation. $\gamma=1/\sqrt[]{1-\beta^2}$ is the relativistic Lorentz factor of the projectile. $\omega_K$ is the K-shell X-ray fluorescence yield, which depends on the charge state of the ion~\cite{Doyle78}. However, the K-shell fluorescence yield only increases by several percent for ions with the electronic configuration $(1s)^1(2s)^2(2p)^5$~\cite{Bhalla73}, compared to neutral atom. The K-shell fluorescence yield for neutral Ni is 0.4~\cite{Kahoul12}. This value has been used in the calculations reported here. A K-shell ionization cross section ($\sigma_{K}$) of 1050 barn for Ni$^{19+}$ ions was adopted to deduce the absolute luminosity in the present work, which was determined with the Relativistic Ionization CODE (RICODE). The RICODE is based on the relativistic Born approximation~\cite{Shevelko11} and is a further development of the LOSS and LOSS-R codes~\cite{Shevelko01}. The RICODE has been widely applied to predict the single-electron loss cross sections for collisions of heavy many-electron ions with neutral atoms in the relativistic energy region.
In this work, a luminosity of 328(6) mb$^{-1}$ was deduced where the error is the statistical uncertainty of the measured X-rays.

\begin{figure}[h!]
\begin{center}
\includegraphics*[width=0.5\textwidth]{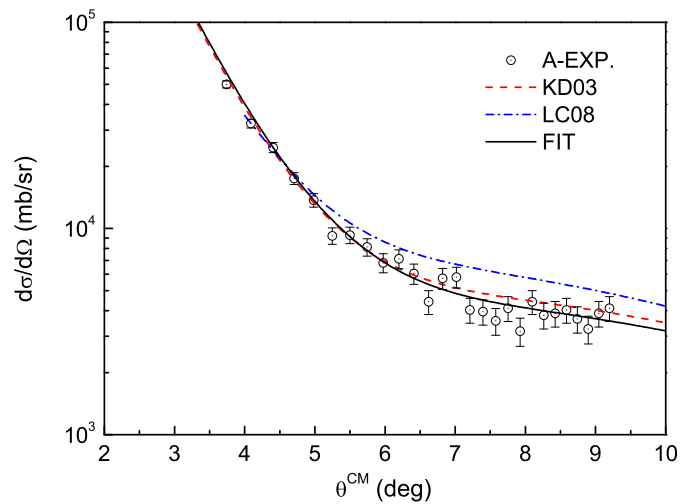}
\caption{(Color online). The measured absolute differential cross sections (A-EXP.) for elastic proton scattering on $^{58}$Ni nuclei as a function of the scattering angle $\theta^{CM}$, which are compared to OMP calculations. A good agreement with KD03 predictions~\cite{Koning03} is observed. The error bars reflect only the statistical uncertainties of the number of the detected protons. The fitting curve is obtained with the SFRESCO code~\cite{Thompson}.}
\label{fig5}
\end{center}
\end{figure}

Compared to the global high accurate OMP predictions, an inconsistency with experimental results was observed for the 100 MeV proton elastic scattering on $^{58}$Ni nuclei in Refs.~\cite{Kwiatkowski78,Koning03}. 
The obtained absolute differential cross sections in the present work as a function of the scattering angle ($\theta^{CM}$) are shown in Fig.~\ref{fig5}, compared to the global OMP results.  A good agreement has been achieved, which proves the reliability of the KD03 calculations~\cite{Koning03} in the measured angle region and clarifies the inconsistency of the cross sections reported in Ref.~\cite{Kwiatkowski78,Koning03}. 

\begin{figure}[h!]
\begin{center}
\includegraphics*[width=0.5\textwidth]{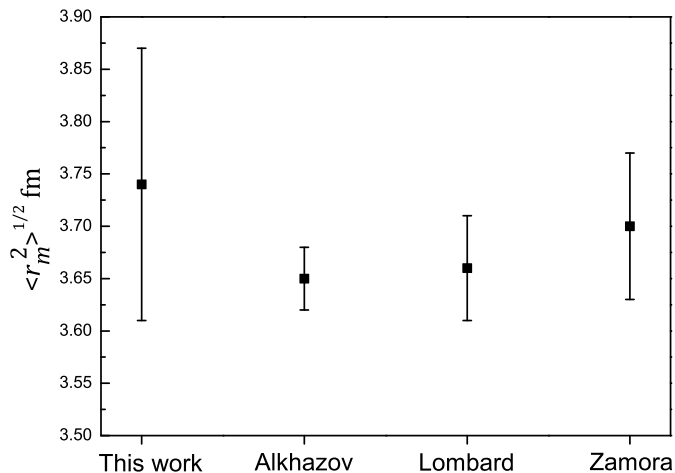}
\caption{The nuclear matter rms value of the $^{58}$Ni nucleus obtained in this work is consistent with data from Zamora et al.~\cite{Zamora17}, Alkhazov et al.~\cite{Alkhazov77} and Lombard et al.~\cite{Lombard81}. The error bar in the present work reflects only the statistical uncertainty.}
\label{fig6}
\end{center}
\end{figure}

The real part of OMPs is related to the nuclear matter distribution~\cite{Vitturi87,Greenlees68}. In the present work, a simple method, suggested by Greenlees et al.~\cite{Greenlees68}, has been used to estimate the nuclear matter rms value of the $^{58}$Ni nucleus.
\begin{equation}\label{e6}
\begin{aligned}
 &\left \langle r^2_m  \right \rangle=\left \langle r^2_{op}  \right \rangle-\left \langle r^2_{2b}  \right \rangle,\\
& \left \langle r^2_{op}  \right \rangle=\frac{3}{5} R^2_v \left[ 1+\frac{7}{3}(\frac{\pi a_v}{R_v})^2 \right],
 \end{aligned}
\end{equation}
where $\left \langle r^2_m  \right \rangle$ is the mean square radius of the nuclear matter distribution, $\left \langle r^2_{2b}  \right \rangle$ is the mean square radius corresponding to the spin and isospin independent part of the two-body potential, an updated value of 4.27 fm$^2$ is adopted in the present analysis, which is obtained by assuming the Gaussian Two-Body Force~\cite{Greenlees70}.
All OMP parameters from KD03~\cite{Koning03} are fixed. Only the nuclear radius ($R_v=r_vA^{1/3}$) and diffuseness ($a_v$) of the real volume potential remain as adjustable parameters to fit the experimental differential cross sections by using the SFRESCO code~\cite{Thompson}. The values $r_v$ and $a_v$ of the real volume potential are determined to be 1.161(15) fm, and 0.667(50) fm, respectively.  Then, the nuclear matter rms value of 3.74(13) fm is obtained via Eq. (\ref{e6}),  which is consistent with the literature results~\cite{Zamora17,Alkhazov77,Lombard81}, see Fig.~\ref{fig6}. Since technically we are able to use exotic nuclei in the same way as described here, our work illustrates the new possibility for performing such studies also on rare systems.

\section{SUMMARY}

The first nuclear reaction experiment with 95 MeV/u $^{58}$Ni$^{19+}$ ions impinging on a hydrogen gas-jet target in a storage ring was successfully performed at the HIRFL-CSR heavy-ion storage ring. A recently developed in-ring experimental method was employed in the experiment.
The low energy protons from the $^{58}$Ni(\emph{p},\emph{p})$^{58}$Ni reaction have been measured to determine the differential cross-sections.
The reaction luminosity was obtained by using the K-shell X-rays from the ionization of $^{58}$Ni$^{19+}$ projectiles by the H$_2$ target, and thus, the absolute differential cross-sections for proton scattering were obtained. 
Our experimental results are in good agreement with KD03 predictions~\cite{Koning03}, which shows the reliability of the KD03 calculations for the measured angular region and clarifies the inconsistency of cross sections reported in the literature~\cite{Kwiatkowski78,Koning03}.
The first successful in-ring reaction experiment demonstrates the applicability of the HIRFL-CSR facility for internal target nuclear reaction studies at the CSRe, and shows a great potential for extracting reliable OMPs of unstable nuclei. A new storage ring complex, the High Intensity heavy ion Accelerator Facility (HIAF)~\cite{Yang13}, will be constructed in China. The first in-ring reaction experiment  at the CSRe is an important step towards the completion of a large angular coverage detection setup intended to be used at a dedicated storage ring of the future HIAF. 

\begin{acknowledgments}
We would like to thank Viatcheslav P. Shevelko for the calculation of the inner-shell ionization cross section.
This work is supported in part by the Major State Basic Research Development Program of China (2016YFA0400503),
by the NSFC Grant No. (11775273, 11575269),
by the CAS Pioneer Hundred Talents Program,
by the CAS Open Research Project of large Research infrastructures,
by the Max-Planck-Society,
by the CAS Maintenance and Reform of large Research infrastructures (DSS-WXGZ-2018-0002),
by the Helmholtz-CAS Joint Research Group HCJRG-108,
by the Deutscher Akademischer Austauschdienst (DAAD) through Programm des Projektbezogenen Personenaustauschs (PPP) with China (57389367),
by the External Cooperation Program of the Chinese Academy of Sciences (GJHZ1305),
and YAL and receives funding from the European Research Council (ERC) under the European Union’s Horizon 2020 research and innovation programme (682841 “ASTRUm”).
\end{acknowledgments}

\bibliography{P+58Ni.bbl}

\begin{thebibliography}{46}%
\makeatletter
\providecommand \@ifxundefined [1]{%
 \@ifx{#1\undefined}
}%
\providecommand \@ifnum [1]{%
 \ifnum #1\expandafter \@firstoftwo
 \else \expandafter \@secondoftwo
 \fi
}%
\providecommand \@ifx [1]{%
 \ifx #1\expandafter \@firstoftwo
 \else \expandafter \@secondoftwo
 \fi
}%
\providecommand \natexlab [1]{#1}%
\providecommand \enquote  [1]{``#1''}%
\providecommand \bibnamefont  [1]{#1}%
\providecommand \bibfnamefont [1]{#1}%
\providecommand \citenamefont [1]{#1}%
\providecommand \href@noop [0]{\@secondoftwo}%
\providecommand \href [0]{\begingroup \@sanitize@url \@href}%
\providecommand \@href[1]{\@@startlink{#1}\@@href}%
\providecommand \@@href[1]{\endgroup#1\@@endlink}%
\providecommand \@sanitize@url [0]{\catcode `\\12\catcode `\$12\catcode
  `\&12\catcode `\#12\catcode `\^12\catcode `\_12\catcode `\%12\relax}%
\providecommand \@@startlink[1]{}%
\providecommand \@@endlink[0]{}%
\providecommand \url  [0]{\begingroup\@sanitize@url \@url }%
\providecommand \@url [1]{\endgroup\@href {#1}{\urlprefix }}%
\providecommand \urlprefix  [0]{URL }%
\providecommand \Eprint [0]{\href }%
\providecommand \doibase [0]{http://dx.doi.org/}%
\providecommand \selectlanguage [0]{\@gobble}%
\providecommand \bibinfo  [0]{\@secondoftwo}%
\providecommand \bibfield  [0]{\@secondoftwo}%
\providecommand \translation [1]{[#1]}%
\providecommand \BibitemOpen [0]{}%
\providecommand \bibitemStop [0]{}%
\providecommand \bibitemNoStop [0]{.\EOS\space}%
\providecommand \EOS [0]{\spacefactor3000\relax}%
\providecommand \BibitemShut  [1]{\csname bibitem#1\endcsname}%
\let\auto@bib@innerbib\@empty
\bibitem [{\citenamefont {Glendenning}(2004)}]{Glendenning04}%
  \BibitemOpen
  \bibfield  {author} {\bibinfo {author} {\bibfnamefont {N.~K.}\ \bibnamefont
  {Glendenning}},\ }\href@noop {} {\emph {\bibinfo {title} {Direct Nuclear
  Reactions}}},\ \bibinfo {number} {ISBN: 978-981-238-945-9}\ (\bibinfo {year}
  {2004})\BibitemShut {NoStop}%
\bibitem [{\citenamefont {Geiger}\ and\ \citenamefont
  {Marsden}(1909)}]{Geiger09}%
  \BibitemOpen
  \bibfield  {author} {\bibinfo {author} {\bibfnamefont {H.}~\bibnamefont
  {Geiger}}\ and\ \bibinfo {author} {\bibfnamefont {E.}~\bibnamefont
  {Marsden}},\ }\href@noop {} {\bibfield  {journal} {\bibinfo  {journal} {Proc.
  R. Soc. Lond. A}\ }\textbf {\bibinfo {volume} {82}},\ \bibinfo {pages} {495}
  (\bibinfo {year} {1909})}\BibitemShut {NoStop}%
\bibitem [{\citenamefont {Egelhof}\ \emph {et~al.}(2002)\citenamefont {Egelhof}
  \emph {et~al.}}]{Egelhof02}%
  \BibitemOpen
  \bibfield  {author} {\bibinfo {author} {\bibfnamefont {P.}~\bibnamefont
  {Egelhof}} \emph {et~al.},\ }\href@noop {} {\bibfield  {journal} {\bibinfo
  {journal} {Eur. Phys. J. A}\ }\textbf {\bibinfo {volume} {15}},\ \bibinfo
  {pages} {27} (\bibinfo {year} {2002})}\BibitemShut {NoStop}%
\bibitem [{\citenamefont {Becchetti}\ and\ \citenamefont
  {Greenlees}(1969)}]{Becchetti69}%
  \BibitemOpen
  \bibfield  {author} {\bibinfo {author} {\bibfnamefont {F.~D.}\ \bibnamefont
  {Becchetti}}\ and\ \bibinfo {author} {\bibfnamefont {G.~W.}\ \bibnamefont
  {Greenlees}},\ }\href@noop {} {\bibfield  {journal} {\bibinfo  {journal}
  {Phys. Rev.}\ }\textbf {\bibinfo {volume} {182}},\ \bibinfo {pages} {1190}
  (\bibinfo {year} {1969})}\BibitemShut {NoStop}%
\bibitem [{\citenamefont {Sakaguchi}\ and\ \citenamefont
  {Zenihiro}(2017)}]{Sakaguchi17}%
  \BibitemOpen
  \bibfield  {author} {\bibinfo {author} {\bibfnamefont {H.}~\bibnamefont
  {Sakaguchi}}\ and\ \bibinfo {author} {\bibfnamefont {J.}~\bibnamefont
  {Zenihiro}},\ }\href@noop {} {\bibfield  {journal} {\bibinfo  {journal}
  {Prog. Part. Nucl. Phys.}\ }\textbf {\bibinfo {volume} {97}},\ \bibinfo
  {pages} {1} (\bibinfo {year} {2017})}\BibitemShut {NoStop}%
\bibitem [{\citenamefont {Koning}\ and\ \citenamefont
  {Delaroche}(2003)}]{Koning03}%
  \BibitemOpen
  \bibfield  {author} {\bibinfo {author} {\bibfnamefont {A.~J.}\ \bibnamefont
  {Koning}}\ and\ \bibinfo {author} {\bibfnamefont {J.~P.}\ \bibnamefont
  {Delaroche}},\ }\href@noop {} {\bibfield  {journal} {\bibinfo  {journal}
  {Nucl. Phys. A}\ }\textbf {\bibinfo {volume} {713}},\ \bibinfo {pages} {231}
  (\bibinfo {year} {2003})}\BibitemShut {NoStop}%
\bibitem [{\citenamefont {Li}\ and\ \citenamefont {Cai}(2008)}]{Li08}%
  \BibitemOpen
  \bibfield  {author} {\bibinfo {author} {\bibfnamefont {X.}~\bibnamefont
  {Li}}\ and\ \bibinfo {author} {\bibfnamefont {C.}~\bibnamefont {Cai}},\
  }\href@noop {} {\bibfield  {journal} {\bibinfo  {journal} {Nucl. Phys. A}\
  }\textbf {\bibinfo {volume} {801}},\ \bibinfo {pages} {43} (\bibinfo {year}
  {2008})}\BibitemShut {NoStop}%
\bibitem [{\citenamefont {Dickhoff}\ and\ \citenamefont
  {Charity}(2019)}]{Dickhoff19}%
  \BibitemOpen
  \bibfield  {author} {\bibinfo {author} {\bibfnamefont {W.~H.}\ \bibnamefont
  {Dickhoff}}\ and\ \bibinfo {author} {\bibfnamefont {R.~J.}\ \bibnamefont
  {Charity}},\ }\href@noop {} {\bibfield  {journal} {\bibinfo  {journal} {Prog.
  Part. Nucl. Phys.}\ }\textbf {\bibinfo {volume} {105}},\ \bibinfo {pages}
  {252} (\bibinfo {year} {2019})}\BibitemShut {NoStop}%
\bibitem [{\citenamefont {Gutbrod}\ \emph {et~al.}(2006)\citenamefont {Gutbrod}
  \emph {et~al.}}]{Gutbrod06}%
  \BibitemOpen
  \bibfield  {author} {\bibinfo {author} {\bibfnamefont {H.~H.}\ \bibnamefont
  {Gutbrod}} \emph {et~al.},\ }\href@noop {} {\emph {\bibinfo {title} {FAIR
  Baseline Technical Report}}},\ \bibinfo {type} {Tech. Rep.}\ \bibinfo
  {number} {ISBN 3-9811298-0-6}\ (\bibinfo {year} {2006})\BibitemShut {NoStop}%
\bibitem [{\citenamefont {Li}\ \emph {et~al.}(2007)\citenamefont {Li} \emph
  {et~al.}}]{Li07}%
  \BibitemOpen
  \bibfield  {author} {\bibinfo {author} {\bibfnamefont {T.}~\bibnamefont {Li}}
  \emph {et~al.},\ }\href@noop {} {\bibfield  {journal} {\bibinfo  {journal}
  {Phys. Rev. Lett.}\ }\textbf {\bibinfo {volume} {99}},\ \bibinfo {pages}
  {162503} (\bibinfo {year} {2007})}\BibitemShut {NoStop}%
\bibitem [{\citenamefont {Chen}\ \emph {et~al.}(2009)\citenamefont {Chen} \emph
  {et~al.}}]{Chen09}%
  \BibitemOpen
  \bibfield  {author} {\bibinfo {author} {\bibfnamefont {X.}~\bibnamefont
  {Chen}} \emph {et~al.},\ }\href@noop {} {\bibfield  {journal} {\bibinfo
  {journal} {Phys. Rev. C}\ }\textbf {\bibinfo {volume} {79}},\ \bibinfo
  {pages} {024320} (\bibinfo {year} {2009})}\BibitemShut {NoStop}%
\bibitem [{\citenamefont {Egelhof}\ \emph {et~al.}(2003)\citenamefont {Egelhof}
  \emph {et~al.}}]{Egelhof03}%
  \BibitemOpen
  \bibfield  {author} {\bibinfo {author} {\bibfnamefont {P.}~\bibnamefont
  {Egelhof}} \emph {et~al.},\ }\href@noop {} {\bibfield  {journal} {\bibinfo
  {journal} {Phys. Scr. T}\ }\textbf {\bibinfo {volume} {104}},\ \bibinfo
  {pages} {151} (\bibinfo {year} {2003})}\BibitemShut {NoStop}%
\bibitem [{\citenamefont {Zamora}\ \emph {et~al.}(2017)\citenamefont {Zamora}
  \emph {et~al.}}]{Zamora17}%
  \BibitemOpen
  \bibfield  {author} {\bibinfo {author} {\bibfnamefont {J.~C.}\ \bibnamefont
  {Zamora}} \emph {et~al.},\ }\href@noop {} {\bibfield  {journal} {\bibinfo
  {journal} {Phys. Rev. C}\ }\textbf {\bibinfo {volume} {96}},\ \bibinfo
  {pages} {034617} (\bibinfo {year} {2017})}\BibitemShut {NoStop}%
\bibitem [{\citenamefont {von Schmid}\ \emph {et~al.}(2015)\citenamefont {von
  Schmid} \emph {et~al.}}]{Schmid15}%
  \BibitemOpen
  \bibfield  {author} {\bibinfo {author} {\bibfnamefont {M.}~\bibnamefont {von
  Schmid}} \emph {et~al.},\ }\href@noop {} {\bibfield  {journal} {\bibinfo
  {journal} {Phys. Scr. T}\ }\textbf {\bibinfo {volume} {166}},\ \bibinfo
  {pages} {014005} (\bibinfo {year} {2015})}\BibitemShut {NoStop}%
\bibitem [{\citenamefont {Egelhof}\ \emph {et~al.}(2015)\citenamefont {Egelhof}
  \emph {et~al.}}]{Egelhof15}%
  \BibitemOpen
  \bibfield  {author} {\bibinfo {author} {\bibfnamefont {P.}~\bibnamefont
  {Egelhof}} \emph {et~al.},\ }\href@noop {} {\bibfield  {journal} {\bibinfo
  {journal} {JPS Conf. Proc.}\ }\textbf {\bibinfo {volume} {6}},\ \bibinfo
  {pages} {020049} (\bibinfo {year} {2015})}\BibitemShut {NoStop}%
\bibitem [{\citenamefont {Zamora}\ \emph {et~al.}(2016)\citenamefont {Zamora}
  \emph {et~al.}}]{Zamora16}%
  \BibitemOpen
  \bibfield  {author} {\bibinfo {author} {\bibfnamefont {J.~C.}\ \bibnamefont
  {Zamora}} \emph {et~al.},\ }\href@noop {} {\bibfield  {journal} {\bibinfo
  {journal} {Phys. Lett. B}\ }\textbf {\bibinfo {volume} {763}},\ \bibinfo
  {pages} {16} (\bibinfo {year} {2016})}\BibitemShut {NoStop}%
\bibitem [{\citenamefont {Mei}\ \emph {et~al.}(2015)\citenamefont {Mei} \emph
  {et~al.}}]{Mei15}%
  \BibitemOpen
  \bibfield  {author} {\bibinfo {author} {\bibfnamefont {B.}~\bibnamefont
  {Mei}} \emph {et~al.},\ }\href@noop {} {\bibfield  {journal} {\bibinfo
  {journal} {Phys. Rev. C}\ }\textbf {\bibinfo {volume} {92}},\ \bibinfo
  {pages} {035803} (\bibinfo {year} {2015})}\BibitemShut {NoStop}%
\bibitem [{\citenamefont {Glorius}\ \emph {et~al.}(2019)\citenamefont {Glorius}
  \emph {et~al.}}]{Glorius19}%
  \BibitemOpen
  \bibfield  {author} {\bibinfo {author} {\bibfnamefont {J.}~\bibnamefont
  {Glorius}} \emph {et~al.},\ }\href@noop {} {\bibfield  {journal} {\bibinfo
  {journal} {Phys. Rev. Lett.}\ }\textbf {\bibinfo {volume} {122}},\ \bibinfo
  {pages} {092701} (\bibinfo {year} {2019})}\BibitemShut {NoStop}%
\bibitem [{\citenamefont {Streicher}\ \emph {et~al.}(2011)\citenamefont
  {Streicher} \emph {et~al.}}]{Streicher11}%
  \BibitemOpen
  \bibfield  {author} {\bibinfo {author} {\bibfnamefont {B.}~\bibnamefont
  {Streicher}} \emph {et~al.},\ }\href@noop {} {\bibfield  {journal} {\bibinfo
  {journal} {Nucl. Instr. and Meth. A}\ }\textbf {\bibinfo {volume} {654}},\
  \bibinfo {pages} {604} (\bibinfo {year} {2011})}\BibitemShut {NoStop}%
\bibitem [{\citenamefont {Mutterer}\ \emph {et~al.}(2015)\citenamefont
  {Mutterer} \emph {et~al.}}]{Mutterer15}%
  \BibitemOpen
  \bibfield  {author} {\bibinfo {author} {\bibfnamefont {M.}~\bibnamefont
  {Mutterer}} \emph {et~al.},\ }\href@noop {} {\bibfield  {journal} {\bibinfo
  {journal} {Phys. Scr. T}\ }\textbf {\bibinfo {volume} {166}},\ \bibinfo
  {pages} {014053} (\bibinfo {year} {2015})}\BibitemShut {NoStop}%
\bibitem [{\citenamefont {Alkhazov}\ \emph {et~al.}(1997)\citenamefont
  {Alkhazov} \emph {et~al.}}]{Alkhazov97}%
  \BibitemOpen
  \bibfield  {author} {\bibinfo {author} {\bibfnamefont {G.~D.}\ \bibnamefont
  {Alkhazov}} \emph {et~al.},\ }\href@noop {} {\bibfield  {journal} {\bibinfo
  {journal} {Phys. Rev. Lett.}\ }\textbf {\bibinfo {volume} {78}},\ \bibinfo
  {pages} {2313} (\bibinfo {year} {1997})}\BibitemShut {NoStop}%
\bibitem [{\citenamefont {Alkhazov}\ \emph {et~al.}(2002)\citenamefont
  {Alkhazov} \emph {et~al.}}]{Alkhazov02}%
  \BibitemOpen
  \bibfield  {author} {\bibinfo {author} {\bibfnamefont {G.~D.}\ \bibnamefont
  {Alkhazov}} \emph {et~al.},\ }\href@noop {} {\bibfield  {journal} {\bibinfo
  {journal} {Nucl. Phys. A}\ }\textbf {\bibinfo {volume} {712}},\ \bibinfo
  {pages} {269} (\bibinfo {year} {2002})}\BibitemShut {NoStop}%
\bibitem [{\citenamefont {Dobrovolsky}\ \emph {et~al.}(2006)\citenamefont
  {Dobrovolsky} \emph {et~al.}}]{Dobrovolsky06}%
  \BibitemOpen
  \bibfield  {author} {\bibinfo {author} {\bibfnamefont {A.~V.}\ \bibnamefont
  {Dobrovolsky}} \emph {et~al.},\ }\href@noop {} {\bibfield  {journal}
  {\bibinfo  {journal} {Nucl. Phys. A}\ }\textbf {\bibinfo {volume} {766}},\
  \bibinfo {pages} {1} (\bibinfo {year} {2006})}\BibitemShut {NoStop}%
\bibitem [{\citenamefont {Xia}\ \emph {et~al.}(2002)\citenamefont {Xia} \emph
  {et~al.}}]{Xia02}%
  \BibitemOpen
  \bibfield  {author} {\bibinfo {author} {\bibfnamefont {J.~W.}\ \bibnamefont
  {Xia}} \emph {et~al.},\ }\href@noop {} {\bibfield  {journal} {\bibinfo
  {journal} {Nucl. Instr. and Meth. A}\ }\textbf {\bibinfo {volume} {488}},\
  \bibinfo {pages} {11} (\bibinfo {year} {2002})}\BibitemShut {NoStop}%
\bibitem [{\citenamefont {Kwiatkowski}\ and\ \citenamefont
  {Wall}(1978)}]{Kwiatkowski78}%
  \BibitemOpen
  \bibfield  {author} {\bibinfo {author} {\bibfnamefont {K.}~\bibnamefont
  {Kwiatkowski}}\ and\ \bibinfo {author} {\bibfnamefont {N.~S.}\ \bibnamefont
  {Wall}},\ }\href@noop {} {\bibfield  {journal} {\bibinfo  {journal} {Nucl.
  Phys. A}\ }\textbf {\bibinfo {volume} {301}},\ \bibinfo {pages} {349}
  (\bibinfo {year} {1978})}\BibitemShut {NoStop}%
\bibitem [{\citenamefont {Tu}\ \emph {et~al.}(2011)\citenamefont {Tu} \emph
  {et~al.}}]{Tu11}%
  \BibitemOpen
  \bibfield  {author} {\bibinfo {author} {\bibfnamefont {X.~L.}\ \bibnamefont
  {Tu}} \emph {et~al.},\ }\href@noop {} {\bibfield  {journal} {\bibinfo
  {journal} {Phys. Rev. Lett.}\ }\textbf {\bibinfo {volume} {106}},\ \bibinfo
  {pages} {112501} (\bibinfo {year} {2011})}\BibitemShut {NoStop}%
\bibitem [{\citenamefont {Zhang}\ \emph {et~al.}(2012)\citenamefont {Zhang}
  \emph {et~al.}}]{Zhang12}%
  \BibitemOpen
  \bibfield  {author} {\bibinfo {author} {\bibfnamefont {Y.~H.}\ \bibnamefont
  {Zhang}} \emph {et~al.},\ }\href@noop {} {\bibfield  {journal} {\bibinfo
  {journal} {Phys. Rev. Lett.}\ }\textbf {\bibinfo {volume} {109}},\ \bibinfo
  {pages} {102501} (\bibinfo {year} {2012})}\BibitemShut {NoStop}%
\bibitem [{\citenamefont {Mao}\ \emph {et~al.}(2016)\citenamefont {Mao} \emph
  {et~al.}}]{Mao16}%
  \BibitemOpen
  \bibfield  {author} {\bibinfo {author} {\bibfnamefont {L.~J.}\ \bibnamefont
  {Mao}} \emph {et~al.},\ }\href@noop {} {\bibfield  {journal} {\bibinfo
  {journal} {Nucl. Instr. and Meth. A}\ }\textbf {\bibinfo {volume} {808}},\
  \bibinfo {pages} {29} (\bibinfo {year} {2016})}\BibitemShut {NoStop}%
\bibitem [{\citenamefont {Shao}\ \emph {et~al.}(2017)\citenamefont {Shao} \emph
  {et~al.}}]{Shao17}%
  \BibitemOpen
  \bibfield  {author} {\bibinfo {author} {\bibfnamefont {C.}~\bibnamefont
  {Shao}} \emph {et~al.},\ }\href@noop {} {\bibfield  {journal} {\bibinfo
  {journal} {Phys. Rev. A}\ }\textbf {\bibinfo {volume} {96}},\ \bibinfo
  {pages} {012708} (\bibinfo {year} {2017})}\BibitemShut {NoStop}%
\bibitem [{\citenamefont {Shao}\ \emph {et~al.}(2013)\citenamefont {Shao} \emph
  {et~al.}}]{Shao13}%
  \BibitemOpen
  \bibfield  {author} {\bibinfo {author} {\bibfnamefont {C.}~\bibnamefont
  {Shao}} \emph {et~al.},\ }\href@noop {} {\bibfield  {journal} {\bibinfo
  {journal} {Nucl. Instr. and Meth. B}\ }\textbf {\bibinfo {volume} {317}},\
  \bibinfo {pages} {617} (\bibinfo {year} {2013})}\BibitemShut {NoStop}%
\bibitem [{\citenamefont {Zhang}\ \emph {et~al.}(2019)\citenamefont {Zhang}
  \emph {et~al.}}]{Zhang19}%
  \BibitemOpen
  \bibfield  {author} {\bibinfo {author} {\bibfnamefont {J.~T.}\ \bibnamefont
  {Zhang}} \emph {et~al.},\ }\href@noop {} {\bibfield  {journal} {\bibinfo
  {journal} {Nucl. Instr. and Meth. A}\ }\textbf {\bibinfo {volume} {948}},\
  \bibinfo {pages} {162848} (\bibinfo {year} {2019})}\BibitemShut {NoStop}%
\bibitem [{\citenamefont {Eichler}\ and\ \citenamefont
  {Stöhlker}(2007)}]{Eichler07}%
  \BibitemOpen
  \bibfield  {author} {\bibinfo {author} {\bibfnamefont {J.}~\bibnamefont
  {Eichler}}\ and\ \bibinfo {author} {\bibfnamefont {T.}~\bibnamefont
  {Stöhlker}},\ }\href@noop {} {\bibfield  {journal} {\bibinfo  {journal}
  {Phys. Rep.}\ }\textbf {\bibinfo {volume} {439}},\ \bibinfo {pages} {1}
  (\bibinfo {year} {2007})}\BibitemShut {NoStop}%
\bibitem [{\citenamefont {Norbury}\ and\ \citenamefont
  {Dick}(2008)}]{Norbury08}%
  \BibitemOpen
  \bibfield  {author} {\bibinfo {author} {\bibfnamefont {J.~W.}\ \bibnamefont
  {Norbury}}\ and\ \bibinfo {author} {\bibfnamefont {F.}~\bibnamefont {Dick}},\
  }\href@noop {} {\bibfield  {journal} {\bibinfo  {journal} {NASA Technical
  Reports Server}\ ,\ \bibinfo {pages} {215543}} (\bibinfo {year}
  {2008})}\BibitemShut {NoStop}%
\bibitem [{\citenamefont {Thompson}(1988)}]{Thompson}%
  \BibitemOpen
  \bibfield  {author} {\bibinfo {author} {\bibfnamefont {I.~J.}\ \bibnamefont
  {Thompson}},\ }\href@noop {} {\bibfield  {journal} {\bibinfo  {journal}
  {Comp. Phys. Rep.}\ }\textbf {\bibinfo {volume} {7}},\ \bibinfo {pages} {167}
  (\bibinfo {year} {1988})}\BibitemShut {NoStop}%
\bibitem [{ORT()}]{ORTEC}%
  \BibitemOpen
  \href@noop {} {\emph {\bibinfo {title} {ORTEC, SLP Series Silicon
  Lithium-Drifted Planar Low-Energy X Ray Detector Product Configuration
  Guide}}}\BibitemShut {NoStop}%
\bibitem [{\citenamefont {Doyle}\ \emph {et~al.}(1978)\citenamefont {Doyle}
  \emph {et~al.}}]{Doyle78}%
  \BibitemOpen
  \bibfield  {author} {\bibinfo {author} {\bibfnamefont {B.~L.}\ \bibnamefont
  {Doyle}} \emph {et~al.},\ }\href@noop {} {\bibfield  {journal} {\bibinfo
  {journal} {Phys. Rev. A}\ }\textbf {\bibinfo {volume} {17}},\ \bibinfo
  {pages} {523} (\bibinfo {year} {1978})}\BibitemShut {NoStop}%
\bibitem [{\citenamefont {Bhalla}(1973)}]{Bhalla73}%
  \BibitemOpen
  \bibfield  {author} {\bibinfo {author} {\bibfnamefont {C.~P.}\ \bibnamefont
  {Bhalla}},\ }\href@noop {} {\bibfield  {journal} {\bibinfo  {journal} {Phys.
  Rev. A}\ }\textbf {\bibinfo {volume} {8}},\ \bibinfo {pages} {2877} (\bibinfo
  {year} {1973})}\BibitemShut {NoStop}%
\bibitem [{\citenamefont {Kahoul}\ \emph {et~al.}(2012)\citenamefont {Kahoul}
  \emph {et~al.}}]{Kahoul12}%
  \BibitemOpen
  \bibfield  {author} {\bibinfo {author} {\bibfnamefont {A.}~\bibnamefont
  {Kahoul}} \emph {et~al.},\ }\href@noop {} {\bibfield  {journal} {\bibinfo
  {journal} {Radiat. Phys. Chem.}\ }\textbf {\bibinfo {volume} {81}},\ \bibinfo
  {pages} {713} (\bibinfo {year} {2012})}\BibitemShut {NoStop}%
\bibitem [{\citenamefont {Shevelko}\ \emph {et~al.}(2011)\citenamefont
  {Shevelko} \emph {et~al.}}]{Shevelko11}%
  \BibitemOpen
  \bibfield  {author} {\bibinfo {author} {\bibfnamefont {V.~P.}\ \bibnamefont
  {Shevelko}} \emph {et~al.},\ }\href@noop {} {\bibfield  {journal} {\bibinfo
  {journal} {Nucl. Instr. and Meth. B}\ }\textbf {\bibinfo {volume} {269}},\
  \bibinfo {pages} {1455} (\bibinfo {year} {2011})}\BibitemShut {NoStop}%
\bibitem [{\citenamefont {Shevelko}\ \emph {et~al.}(2001)\citenamefont
  {Shevelko} \emph {et~al.}}]{Shevelko01}%
  \BibitemOpen
  \bibfield  {author} {\bibinfo {author} {\bibfnamefont {V.~P.}\ \bibnamefont
  {Shevelko}} \emph {et~al.},\ }\href@noop {} {\bibfield  {journal} {\bibinfo
  {journal} {Nucl. Instr. and Meth. B}\ }\textbf {\bibinfo {volume} {184}},\
  \bibinfo {pages} {295} (\bibinfo {year} {2001})}\BibitemShut {NoStop}%
\bibitem [{\citenamefont {Vitturi}\ and\ \citenamefont
  {Zardi}(1987)}]{Vitturi87}%
  \BibitemOpen
  \bibfield  {author} {\bibinfo {author} {\bibfnamefont {A.}~\bibnamefont
  {Vitturi}}\ and\ \bibinfo {author} {\bibfnamefont {F.}~\bibnamefont
  {Zardi}},\ }\href@noop {} {\bibfield  {journal} {\bibinfo  {journal} {Phys.
  Rev. C}\ }\textbf {\bibinfo {volume} {36}},\ \bibinfo {pages} {1404}
  (\bibinfo {year} {1987})}\BibitemShut {NoStop}%
\bibitem [{\citenamefont {Greenlees}\ \emph {et~al.}(1968)\citenamefont
  {Greenlees} \emph {et~al.}}]{Greenlees68}%
  \BibitemOpen
  \bibfield  {author} {\bibinfo {author} {\bibfnamefont {G.~W.}\ \bibnamefont
  {Greenlees}} \emph {et~al.},\ }\href@noop {} {\bibfield  {journal} {\bibinfo
  {journal} {Phys. Rev.}\ }\textbf {\bibinfo {volume} {171}},\ \bibinfo {pages}
  {1115} (\bibinfo {year} {1968})}\BibitemShut {NoStop}%
\bibitem [{\citenamefont {Greenlees}\ \emph {et~al.}(1970)\citenamefont
  {Greenlees} \emph {et~al.}}]{Greenlees70}%
  \BibitemOpen
  \bibfield  {author} {\bibinfo {author} {\bibfnamefont {G.~W.}\ \bibnamefont
  {Greenlees}} \emph {et~al.},\ }\href@noop {} {\bibfield  {journal} {\bibinfo
  {journal} {Phys. Rev. C}\ }\textbf {\bibinfo {volume} {1}},\ \bibinfo {pages}
  {1145} (\bibinfo {year} {1970})}\BibitemShut {NoStop}%
\bibitem [{\citenamefont {Alkhazov}\ \emph {et~al.}(1977)\citenamefont
  {Alkhazov} \emph {et~al.}}]{Alkhazov77}%
  \BibitemOpen
  \bibfield  {author} {\bibinfo {author} {\bibfnamefont {G.}~\bibnamefont
  {Alkhazov}} \emph {et~al.},\ }\href@noop {} {\bibfield  {journal} {\bibinfo
  {journal} {Phys. Lett. B}\ }\textbf {\bibinfo {volume} {67}},\ \bibinfo
  {pages} {402} (\bibinfo {year} {1977})}\BibitemShut {NoStop}%
\bibitem [{\citenamefont {Lombard}\ \emph {et~al.}(1981)\citenamefont {Lombard}
  \emph {et~al.}}]{Lombard81}%
  \BibitemOpen
  \bibfield  {author} {\bibinfo {author} {\bibfnamefont {R.~M.}\ \bibnamefont
  {Lombard}} \emph {et~al.},\ }\href@noop {} {\bibfield  {journal} {\bibinfo
  {journal} {Nucl. Phys. A}\ }\textbf {\bibinfo {volume} {360}},\ \bibinfo
  {pages} {233} (\bibinfo {year} {1981})}\BibitemShut {NoStop}%
\bibitem [{\citenamefont {Yang}\ \emph {et~al.}(2013)\citenamefont {Yang} \emph
  {et~al.}}]{Yang13}%
  \BibitemOpen
  \bibfield  {author} {\bibinfo {author} {\bibfnamefont {J.~C.}\ \bibnamefont
  {Yang}} \emph {et~al.},\ }\href@noop {} {\bibfield  {journal} {\bibinfo
  {journal} {Nucl. Instr. and Meth. B}\ }\textbf {\bibinfo {volume} {317}},\
  \bibinfo {pages} {263} (\bibinfo {year} {2013})}\BibitemShut {NoStop}%
\end{thebibliography}%

\end{document}